\documentclass[11pt]{article}
\usepackage{graphicx}
\renewcommand{\baselinestretch}{1}
\begin{document}
\title{\bf  Multipole characteristics of \\ the open-shell electron eigenstates}
\author{\bf J. Mulak$^{1}$, M. Mulak$^{2}$}
\date{{\it  $^{1}$ Trzebiatowski Institute of Low Temperature
            and Structure Research,\\
            Polish Academy of Sciences, 50--950, PO Box 1410,
            Wroclaw, Poland\\
            $^{2}$ Institute of Physics,
            Wroclaw University of Technology,\\
            Wyb. Wyspianskiego 27,
            50--370 Wroclaw, Poland}}
\maketitle
\begin{abstract}
\noindent The second moment of the sublevels within the initial state $|\alpha SLJ\rangle$ which constitutes a
natural and adequate measure of the crystal-field (CF) effect can be redefined as
$\sigma^{2}=\frac{1}{2J+1}\sum_{k}S_{k}^2A_{k}^{2}$, where
$S_{k}=\left(\frac{1}{2k+1}\sum_{q}|B_{kq}|^2\right)^{1/2}$ is the so-called $2^{k}$-pole CF strength, whereas
$A_{k}=\langle \alpha SLJ||C^{(k)}||\alpha SLJ\rangle$ the reduced matrix element of the $k$-rank spherical
tensor operator. Therefore, the CF effect depends on the sum of products of the two factors representing the
identical multipole components of two different charge distributions. The term $A_{k}$ expresses the asphericity
of the central ion open-shell, whereas the term $S_{k}$ the asphericity of its surroundings. When these two
distributions do not fit each other the observed CF splitting can be unexpectedly weak even for considerable
values of the total $S=\left(\sum_{k}S_{k}^{2}\right)^{1/2}$ and $A=\left(\sum_{k}A_{k}^{2}\right)^{1/2}$. The
tabulated quantities of the $A_{k}\left(|\alpha SLJ\rangle\right)$, as the $2^{k}$-pole type asphericities, are
the intrinsic characteristics of the electron states revealing their multipolar structure and hence their
potential susceptibility to CF splitting separately for each effective multipole. For any chosen pair of a
central-ion and CF potential the relevant $A_{k}$ and $S_{k}$ magnitudes, respectively, allow us to predict the
scale of the related splitting. We can also compare the CF splitting of various states in the same CF potentials
or the splitting of the same state in various CF potentials. Having the model $\sigma^{2}$ and their
experimentally available counterparts we can evaluate the degree of admixing of the free-ion states. Since the
independent quantities $S_{k}$ and $A_{k}$ occur as the scalar product in the formula for $\sigma^{2}$ the use
of the total $S$ and $A$ notions should be critically considered.
\end{abstract}

\section*{1. Introduction}
The origin of the one-electron parametric CF Hamiltonian ${\cal H}_{\rm CF}=\sum_{k,q}B_{kq}C_{q}^{(k)}$ [1-3],
stemming from the spherical harmonic addition theorem [4-6], shows that it is a sum of products of two terms.
They represent the identical multipole components of two different and independent charge distributions. The
first term, expressed by the set of CF parameters (CFPs) $B_{kq}$, corresponds to the multipole distribution of
the central ion surroundings potential. The second term $C_{q}^{(k)}$ refers exclusively to the multipole
expansion of the angle parts of electron wave functions of the central ion open-shell. This inherent dichotomy
is even more clear, independently of the reference frame, for the second moment of the sublevels within the
initial state $|\alpha, SLJ\rangle$ [7-9]
\begin{equation}
\sigma^{2}\left(|\alpha SLJ\rangle\right)=\frac{1}{2J+1}\sum\limits_{k=2,4,6}S_{k}^{2} \left(\langle\alpha
SLJ||C^{(k)}||\alpha SLJ\rangle\right)^{2}.
\end{equation}
Here, the central-ion surroundings is represented by the $2^{k}$-pole CF strength parameters \linebreak
$S_{k}=\left(\frac{1}{2k+1}\sum_{q}|B_{kq}|^2\right)^{1/2}$ [10-14]. In turn, the squares of the reduced matrix
elements of the $k$-rank spherical tensor operator $(k=2,4,6)$ [15-17] characterize the multipole structure of
the central ion open-shell in the state $|\alpha SLJ\rangle$ and give a measure of its latent asphericity. The
latter quantities determine the state susceptibility to CF splitting induced by the particular $2^{k}$-poles of
the ${\cal H}_{\rm CF}$. The symbol $\alpha$ in Eq.(1) stands for the remaining eigenvalues of the operators
which are invariant with respect to the rotations and define the state $|\alpha SLJ\rangle$.

In the paper we present a thorough analysis of the reduced matrix elements of the spherical tensor operators
$A_{k}=\langle \alpha SLJ||C^{(k)}||\alpha SLJ\rangle$, $k=2,4,6$. The $A_{k}$s for all the states coming from
the ground terms of the basic electron configurations $p^{n}$, $d^{n}$ and $f^{n}$ are enclosed in Table 1.
Additionally, the multipolar characteristics of many other interesting states chosen according to various
criteria are comprised in Table 2. Among them there are the states of the extreme asphericity, those having the
uniform or differentiated asphericity with respect to $k$, as well as the states of the repeated terms (with the
same $L$ and $S$). The last column in the tables gives the total asphericity  of the states
$A=\left(\sum_{k}A_{k}^{2}\right)^{1/2}$, which is a counterpart of the total CF strength
$S=\left(\sum_{k}S_{k}^{2}\right)^{1/2}$, used widely so far [10-14]. The complete database enclosing 283 ionic
(atomic) terms and all the 923 $|\alpha SLJ\rangle$ states for $p^{n} (n=1,2)$, $d^{n} (n=1,2,3,4)$ and $f^{n}
(n=1,2,3,4,5,6)$ configurations will be presented separately. The $\langle \alpha SLJ||C^{(k)}||\alpha
SLJ\rangle$ matrix elements for the remaining $l^{2(2l+1)-n}$ configurations are identical with those for to the
$l^{n}$ ones. The states derived from the half-filled electron configurations $l^{2l+1}$ do not yield any
crystal-field splitting in the first order approximation [18], i.e. all the relevant diagonal reduced matrix
elements are equal to zero. The asphericity $A_{k}$ changes from state to state and any correlation seems to be
difficult to be found. Nevertheless, these seemingly random variations can be interpreted using the reduced
matrix element factorization. Based on the tabulated characteristics of the initial states, even under
limitations to the pure Russell-Saunders coupling, one can predict which CF potentials are effective for a given
state or which states are sensitive to a given CF potential. The analyzed asphericities can also be helpful to
verify any CF data.

\section*{2. General formulation}
We start with the one-electron parametric form of the crystal-field Hamiltonian in the tensor notation by
Wybourne [1]
\begin{equation}
{\cal H}_{\rm CF}=\sum_{i}\sum_{k}\sum_{q}B_{kq}C_{q}^{(k)}(\vartheta_{i},\varphi_{i}),
\end{equation}
where $i$ runs over the central-ion open-shell electrons, precisely over their angle spherical coordinates,
whereas $k$ and $q$ over all the effective multipole components of the CF potential.

In general, this one-electron approach describes correctly most of the problems involving the CF effect with
satisfying accuracy. The best evidence of it is the long list of successful interpretations of various
experimental results [19-22]. However, the origin of the ${\cal H}_{\rm CF}$ formula Eq.(2) descending from the
spherical harmonic addition theorem [4-6] allows us to interpret this expression in the spirit of the complete
partition of the system into two separate multipolar subsystems -- the central ion, and its surroundings
(compare the expansion of $1/r_{ij}$). Keeping up the generality of the considerations from the viewpoint of the
one-electron ${\cal H}_{\rm CF}$ parametrization we may confine the problem to the electrostatic interaction of
the individual ligand of charge $q$ and radius vector ${\bf R}(R,\Theta,\Phi)$ with an individual open-shell
electron of the radius vector ${\bf r}(r,\vartheta,\phi)$. Then, for $R>r$ [2,3]
\begin{equation}
{\cal H}_{\rm CF}=\frac{q\cdot e}{|{\bf R}-{\bf r}|}=q\cdot e \sum_{k}\sum_{q}\frac{r^{k}}{R^{k+1}}C_{q}^{(k
)^{\ast}}(\Theta,\Phi)\cdot C_{q}^{(k )}(\vartheta,\varphi).
\end{equation}
In fact, the ${\cal H}_{\rm CF}$ is the sum of the binary products of the terms representing the identical
multipole components $(k,q)$ of the independent charge distributions in the common central-ion reference frame.
The set of CFPs $B_{kq}=q\cdot e \left(r^{k}/R^{k+1}\right)C_{q}^{(k )^{\ast}}(\Theta,\Phi)$ refers to the
multipolar distribution of the surroundings potential. The averaged $\langle r^{k}\rangle$ over the radial
distribution of the central-ion open-shell electron wave function commonly included into the $B_{kq}$ does not
violate the dychotomic separation [1,3,6]. In turn, the second distribution represented by $C_{q}^{(k
)}(\vartheta,\varphi)$ concerns exclusively the angle distribution of the open-shell electron density of the
unperturbed central ion and constitues its intrinsic characteristics. The crucial point is that only the
products of the identical complex-conjugate multipole components of both the distributions, in other words their
scalar products, contribute to the ${\cal H}_{\rm CF}$. On the contrary, the orthogonal multipolar distributions
of the central-ion electronic state and the surroundings potential do not contribute to the CF interaction.

\subsection*{2.1. Factorization of the ${\cal H}_{\rm CF}$ matrix elements}
The crystal-field impact on the $(2J+1)$-fold degenerate electron state $|\alpha SLJ\rangle$ is determined by
the ${\cal H}_{\rm CF}$ matrix elements $\langle \alpha SLJM_{J}|{\cal H}_{\rm CF}|\alpha
SLJM_{J}^{\prime}\rangle$ within the state. This statement holds true exclusively within the limits of the first
order perturbational approach, it means taking into account only the diagonal matrix elements with respect to
$J$ . Then, to describe the crystal field splitting of the $|\alpha SLJ\rangle$ state it is sufficient to know
its multipole characteristics and the relevant set of the CFPs. In the case of more complex initial eigenstates
formed in various types of mixing mechanisms (higher order perturbations, simultaneous diagonalization) the
multipolar characteristics are available including all the needed off-diagonal matrix elements of the operators
$C^{(k)}$ or $U^{(k)}$ [18].

The ${\cal H}_{\rm CF}$ matrix elements undergo a far-reaching factorization which provides understanding of
their seemingly random variations from state to state arising simply from their origin. Using the Wigner-Eckart
theorem in the first step of this factorization one gets [4,15-17,23]
\begin{equation}
\langle \alpha SLJM_{J}|B_{kq}C_{q}^{(k)}|\alpha SLJM_{J}^{\prime}\rangle=(-1)^{J-M_{J}}B_{kq}
\left(\begin{array}{ccc}
J&k&J\\
-M_{J}&q&M_{J^{\prime}} \end{array} \right)\langle \alpha SLJ ||C^{(k)}||\alpha SLJ\rangle.
\end{equation}
The first three factors on the right hand, i.e. the one defining the sign, the respective $B_{kq}$ CFP, and the
$3-j$ symbol, determine the splitting pattern of the $|\alpha SLJ\rangle$ state in the assumed reference system.
The last dimensionless term, the so-called double bar or reduced matrix element, independent of the reference
frame,  is the factor scaling the splitting. The factorization process can be continued further and for the
particular case of the diagonal matrix elements it takes the form [4,15-17,23]
\begin{eqnarray}
  \langle\alpha SLJ||C^{(k)}||\alpha SLJ\rangle &=&
  (-1)^{S+L+J+k}
  (2J+1)\left\{\begin{array}{ccc}
                 J & J & k \\
                 L & L & S \\
                \end{array}\right\}
\langle\alpha SL||C^{(k)}||\alpha SL\rangle,
\end{eqnarray}
where the doubly reduced element (the last one independent on the J quantum number) is usually written down by
means of the unit tensor operator $U^{(k)}$ as [18,23]
\begin{equation}
\langle \alpha SL ||C^{(k)}||\alpha SL\rangle=\langle \alpha SL ||U^{(k)}||\alpha SL\rangle \langle  l
||C^{(k)}|| l\rangle,
\end{equation}
where $l=1,2$ and $3$ for $p$, $d$ and $f$ electrons, respectively. The first factor in Eq.(5) defining the sign
of the reduced element depends on the parity of the sum of four relatively autonomous numbers what undoubtedly
leads to its random character. The $6-j$ symbol multiplied by the $(2J+1)$- degeneracy degree of the state
reveals, according to its physical meaning, what part of the final $|\alpha SLJ\rangle$ function belongs to the
orbital part $|\alpha SL\rangle$. Since the ${\cal H}_{\rm CF}$ acts exclusively on the configurational
coordinates of the electrons only this part is responsible for the CF interaction with the $2^{k}$-pole
component of the CF potential. The expressions $(2J+1)\left\{\begin{array}{ccc}
J & J & k \\
L & L & S \\
\end{array}\right\}$ have their own extra sign, and their magnitudes span from $0$ to $1$. Only in
certain cases for $k=2$ they exceed $1$ reaching the largest value of $1.1096$ for the state $^{7}F_{6}(f^{6})$
[23,24]. Finally, the doubly reduced matrix element depends on the way of coupling of the $n$ one-electron
angular momentum ${\bf l}$ of the $l^{n}$ configuration to the resultant ${\bf L}$, since each one-electron
component state $|lm\rangle$ reacts individually to the outer CF potential according to its orientation in the
final reference frame. There are as many different doubly reduced matrix elements having the same $L$ and $S$
numbers as the different ways of the coupling. In the group-theory language -- there are as many different terms
of the same $L$ and $S$ as the irreducible representations ${\cal D}^{(L)}$ of the three-dimensional rotations
group in the decomposition of the simple Kronecker product ${\cal D}^{(l)}(1)\times {\cal D}^{(l)}(2) \ldots
{\cal D}^{(l)}(n)$ regarding however the Pauli exclusion principle. This differentiation of the states arising
from their genealogy can be expressed equivalently by the Racah fractional parentage coefficients [1,17] or by
the seniority numbers [1,5]. The doubly reduced matrix elements $\langle \alpha SL ||C^{(k)}||\alpha SL\rangle$
change from $-1.8899$ for the term $^{1}I(d^{4})$ and $k=4$ up to $2.5803$ for the term $^{1}N(f^{4})$ and
$k=2$. The absolute values of the one-electron reduced matrix elements $\langle  l ||C^{(k)}|| l\rangle$ differ
only weakly from $1.0955$ to $1.3663$ for all the $k=1,2$ and $3$ and $l=1,2$ and $3$.

Additionally, there exists a one-to-one correspondence between the reduced matrix elements \linebreak
$\langle\alpha SLJ||C^{(k)}||\alpha SLJ\rangle$ and the multiplicative Stevens coefficients $\alpha, \beta,
\gamma$ for $k=2,4,6$, in his operator equivalent method [25,26]. In fact, the equivalent operator matrix
elements $\langle J M_{J} | \hat{O}_{k}^{q} |JM_{J}^{\prime}\rangle$ are equal to the $3-j$ symbols
$\left(\begin{array}{ccc}
J & k & J \\
-M_{J} & q & M_{J}^{\prime} \\
\end{array}\right)$ with the accuracy to the constant coefficient.

\subsection*{2.2. The second moment of the sublevels within the initial state $|\alpha SLJ\rangle$}
The second moment $\sigma^{2}\left(|\alpha SLJ\rangle\right)$ of the sublevels within the initial state $|\alpha
SLJ\rangle$ [7-9] (Eq.(1)) is a natural and adequate measure of the CF effect. It is an important invariant of
the three-dimensional rotation group $R_{3}$. This property implies from the orthogonality of the $3-j$ symbols
[11,23]. In consequence, $\sigma^{2}$ is not literally dependent on the CFPs determining the splitting in any
definite reference system but instead it depends on the invariant $S_{k}$. As a result the contributions of the
individual multipoles to the $\sigma^{2}$ are additive.

To simplify the notation let us introduce
\begin{equation}
A_{k}(|\alpha SLJ\rangle)=\langle\alpha SLJ||C^{(k)}||\alpha SLJ\rangle.
\end{equation}
This dimensionless scalar depending only on the angle distribution of the electron density of the state $|\alpha
SLJ\rangle$ reflects its latent asphericity of the $2^{k}$-pole type. Now,
\begin{equation}
\sigma^{2}\left(|\alpha SLJ\rangle\right)=\frac{1}{2J+1}\sum\limits_{k=2,4,6}S_{k}^{2} A_{k}^{2},
\end{equation}
which means that the CF effect is not only the function of the $S_{k}$ and $A_{k}$ magnitudes but primarily
depends on the degree of overlapping of these two separate multipole distributions that characterize
respectively the surroundings and the central ion. By analogy with the total CF strength notion
$S=(S_{2}^{2}+S_{4}^{2}+S_{6}^{2})^{1/2}$ which refers to the asphericity of the surroundings we introduce the
concept of the total asphericity of the central ion state $|\alpha SLJ\rangle$
\begin{equation}
A=(A_{2}^{2}+A_{4}^{2}+A_{6}^{2})^{1/2}.
\end{equation}
It bears the same advantages and faults as the total CF strength $S$. Since the $A_{k}$ are dimensionless the
energy dimension of the $\sigma$ in Eq.(8) is ensured by the $S$, i.e. by the CFPs unit.

The total asphericity $A$ is $0$ for all the terms coming from the half-filled configurations $l^{2l+1}$ and
terms $S$ having $L=0$. From among all the considered free-ion states the maximal value of $A=4.2225$ (Table 2).

\section*{3. Review of the multipole characteristics of the initial eigenstates. Calculational results}
The multipolar characteristics of the pure free-ion states $|\alpha SLJ\rangle$ with respect to the $2^{2}$-,
$2^{4}$- and $2^{6}$-multipoles active in the CF interaction are reviewed thoroughly below. From among all the
923 essentially various states of the simple $p^{n}$, $d^{n}$ and $f^{n}$ electron configurations only 121 of
them have been chosen for the presentation and analysis. Of necessity our discussion is here restricted only to
this survey. The selected states are either the most important or the most distinguishable ones and properly
illustrate their random-like variability. The reduced matrix elements of the spherical tensor operators $A_{k}$,
introduced in the paper as the $2^{k}$-pole type asphericities which represent the multipolar character of the
states as well as the total asphericity $A=\left(\sum_{k}A_{k}^{2}\right)^{1/2}$ tentatively revealing the
susceptibility of the states to CF splitting are compiled in Tables 1 and 2. The multipole components $A_{k}$
have their own sign (Eq.(5)) which determines the sign of the energy eigenvalues, but is inessential in the
second moment (Eqs (1,8)). Tables 1 and 2 contain also the $A_{k}$ and $A$ calculated for all the states
$^{2S+1}L$ ignoring the spin-orbit coupling no matter how strong it is in relation to the CF interaction. The
data for the ground terms of all the electron configurations are given in Table 1, whereas the data for the
selected states are compiled in Table 2. In the latter one can find the multipolar characteristics of the states
of the strongest or weakest susceptibility to CF splitting, the states selectively sensitive to the $2^{2}$-,
$2^{4}$- and $2^{6}$-pole components of the ${\cal H}_{\rm CF}$, as well as sensitive to the three multipoles in
a comparable measure, and finally the states coming from the repeated terms, i.e. of the same $L$ and $S$
numbers. The maximal global asphericity $A=4.2225$ has been found for the state $^{1}N(f^{4})$, whereas the
highest asphericities for the respective multipoles amount to: $|A_{2}|=3.5254$, $|A_{4}|=2.2589$ and
$|A_{6}|=2.7260$ correspondingly for the states $^{1}N(f^{4})$, $^{1}I(d^{4})$ and $^{1}Q(f^{6})$. The common
feature of these top-level states is the largest possible $L$ numbers $(L=10,6,12$ respectively) and $S=0$. The
$|\overline{A_{k}}|$ and $|\overline{A}|$ asphericities averaged over all the 923 states amount to:
$|\overline{A_{2}}|=0.5244$, $|\overline{A_{4}}|=0.3222$, $|\overline{A_{6}}|=0.2904$ and
$|\overline{A}|=0.7967$. The state $^{4}I_{11/2}(f^{3})$ (Table 1) well represents the average susceptibility to
CF splitting, and therefore can serve as the reference state.

The reduced matrix elements $A_{k}=\langle J||C^{(k)}|| J\rangle$ are composed of the five factors. This
complexity is responsible for a seemingly stochastic distribution of $A_{k}$ magnitudes and signs (subsection
2.1). The matrix elements for various $k$ (the rows in Tables) are completely independent from each other as
they belong to various multipoles, it means to the various irreducible representations of the $R_{3}$ group. For
the fixed $k$ (the columns in Tables) we can observe within the multiplets $^{2S+1}L$ certain systematic
correlation between the $A_{k}$ and $J$ values of the relevant states. The largest absolute value of the
$\langle J||C^{(k)}|| J\rangle$ element is reached for the state with $J=L+S$ and it is close, as a rule, to the
value calculated ignoring the spin. For the smaller $J$ the corresponding $A_{k}$s decrease to rise for the
smallest $J$ number within the multiplet. This systematic correlation results from the ratio (Eq.(5)) and takes
the form

\begin{equation}
\frac{\langle J||C^{(k)}|| J\rangle}{\langle J+n||C^{(k)}|| J+n\rangle}= (-1)^{n} \frac{(2J+1)}{(2J+2n+1)}
\frac{ \left(\begin{array}{ccc}
J&J&k\\
L&L&S \end{array} \right)}{\left(\begin{array}{ccc}
J+n&J+n&k\\
L&L&S \end{array} \right)}
\end{equation}
Figuratively, such a behaviour can be explained by the degree of the $\bf{L}$ and $\bf{J}$ vectors co-linearity.
Based on the above relation we can find the ratios of the individual multipoles contributions to the second
moment $\sigma^{2}$ of the sublevels within the initial states $|\alpha SLJ\rangle$ of the same term $|\alpha
SL\rangle$.

\section*{4. Discussion}
The distinction between the multipole characteristics of the states originating from the ground terms of the
$p^{n}$, $d^{n}$ and $f^{n}$ electron configurations is shown in Table 1. The observed dispersion of the $A_{k}$
and $A$ values confronted with the multipole characteristics of the involved CF potential provides understanding
of the role of the individual multipoles of both the distributions in the formation of the actual splitting
pattern.

As an example let us consider the $f^{2}$ configuration occurring in Pr$^{3+}$ and U$^{4+}$ ions. The states of
the $^{3}H$ term, in particular the $^{3}H_{6}$ one, are characterized by strong or averaged asphericity of all
the three multipole types, whereas the states of the $^{3}F$ term by average or weak asphericity. The $^{1}I$
state is exceptionally likely to be affected by the quadrupolar $(k=2)$ component of the ${\cal H}_{\rm CF}$,
while the $^{1}G$ state by the higher order multipoles $(k=4,6)$. Based on these data we can predict the states
splittings in crystal fields of different point symmetries or to verify the proportions between the multipole
terms in the involved ${\cal H}_{\rm CF}$. In turn, the states $^{4}I_{15/2}$, $^{4}G_{11/2}$ and $^{4}F_{9/2}$
of the $f^{3}$ electron configuration are distinguished by the prevailing role of the $2^{6}$-, $2^{4}$- and
$2^{2}$-pole component, respectively. This asphericity hierarchy of the states governs their splitting when the
states are subjected to the particular ${\cal H}_{\rm CF}$.

Some selected $|\alpha SLJ\rangle$ states which substantially differ in their multipolar characteristics are
gathered in Table 2. However, due to the complex structure of the reduced matrix elements (subsection 2.1), only
a cursory interpretation of the characteristics in terms of the intrinsic quantum numbers of the states is
feasible. Specifically, the states of the highest total asphericity $(A>2.5)$  stand out by the large quantum
numbers $L$ and small $S$ (mostly $S=0$). The states of the lowest asphericity $(A<0.2)$ are derived rather from
the almost half-filled configurations and are characterized by average or small $L$ and irregular $S$ numbers
(typically $S=1$). In turn, the states with the maximal $A_{2}$ asphericity show the same features as those of
the high total $A$. In case of the states with the highest $A_{4}$ asphericity no particular preferences for the
$L$ and $S$ numbers are observed, although several of them descend from the $d$-electron configurations.
Finally, the states with the highest $A_{6}$ asphericity arise rather from the almost half-filled configurations
and they are characterized as a rule by large $L$ numbers. The states of the selectively dominating component
$A_{k}$ (respective to the remaining two $A_{k}$s) are presented in rows 21--44 of Table 2. Among the states of
the comparable contributions of the three multipoles intriguingly many of them come from the $F$ terms. To have
a detailed insight into the $A_{k}$ and $A$ distributions and their interpretation one has to carefully study
the complete data set compiling the multipolar characteristics of all the states. The seemingly stochastic
distribution of the $A_{k}$ and $A$ values (Tables 1 and 2) is well demonstrated by the chosen seven various
$^{3}G_{5}$ states of $f^{6}$ or $f^{8}$ configurations. In this case the ratios between the seven relevant
$A_{k}$s are equal to the ratios between the seven corresponding reduced matrix elements $\langle \alpha SL
||U^{(k)}||\alpha SL\rangle$ with various $\alpha$. The formula for the $\sigma^{2}$ (Eq.(8)) gives the direct
relationship (separately for the $2^{k}$-pole contributions) between the total second moment representing the
spectrum of the energy levels from the one hand, and both the $S_{k}$ (dependent on the $B_{kq}$ CFPs) and the
$A_{k}$ (the $2^{k}$-pole characteristics of the $|\alpha SLJ\rangle$ state) from the second hand. The
relationship is sensitive to variations of the composing terms. It provides a convenient test for the
correctness of the initial assumptions in the CFPs fitting procedure since the number of the involved variables
is now considerable reduced. Applying the formula on the $\sigma^{2}$ we are able to narrow the set of possible
splitting patterns and more easily localize the states being invisible in the spectroscopy. It allows us also to
estimate the degree of admixture of the initial states comparing the actual $A_{k}$s with their free-ion
counterparts.

Here, as an example, let us consider the well evidenced crystal-field splitting of the $^{3}H_{4}$ and
$^{3}H_{5}$ states in the Pr$^{3+}(f^{2})$ ion in LaF$_{3}$ matrix of the $C_{2}$ point symmetry [27].
Unfortunately, the experimental energies of all the thirteen CF sublevels of the $^{3}H_{6}$ states in this
field remain unknown. From the spectroscopic data [27] result the following second moments: $\sigma_{\rm
exp}^{2}(^{3}H_{4})=22103\; (cm^{-1})^{2}$ and $\sigma_{\rm exp}^{2}(^{3}H_{5})=12285\;(cm^{-1})^{2}$. Based on
the CFPs derived from a fit to the experimental data using a $C_{2v}$ approximation of the actual $C_{2}$
symmetry [27] we get: $S_{2}^{2}=13105\;(cm^{-1})^{2}$, $S_{4}^{2}=152758\;(cm^{-1})^{2}$ and
$S_{6}^{2}=272240\;(cm^{-1})^{2}$. Substituting now the tabulated values (Table 1): $A_{2}=-1.2367$,
$A_{4}=0.7395$ and $A_{6}=0.7706$ for the $^{3}H_{4}$ state and $A_{2}=-1.3100$, $A_{4}=0.6833$ and
$A_{6}=0.4451$ for the $^{3}H_{5}$ state respectively, one obtains $\sigma_{\rm
calc}^{2}(^{3}H_{4})=29474\;(cm^{-1})^{2}$ and $\sigma_{\rm calc}^{2}(^{3}H_{5})=16416\;(cm^{-1})^{2}$. The
$\sigma_{\rm calc}^{2}$ calculated for the model $A_{k}$s exceed the experimental $\sigma_{\rm exp}^{2}$ by
30\%, but interestingly their ratios remain exceptionally close to each other (Eq.(10))
\begin{eqnarray*}
\frac{\sigma_{\rm exp}^{2}(^{3}H_{4})}{\sigma_{\rm exp}^{2}(^{3}H_{5})}=1.799\qquad\mathrm{and}\qquad
\frac{\sigma_{\rm calc}^{2}(^{3}H_{4})}{\sigma_{\rm calc}^{2}(^{3}H_{5})}=1.795.
\end{eqnarray*}
It may be well to add that the ratios of the individual $2^{k}$-poles contributions to the total second moment
are roughly $1:4:8$ and $1:3:2$ for the $^{3}H_{4}$ and $^{3}H_{5}$ states and $k=2,4,6$, respectively. It means
that the decrease of the $A_{6}$ asphericity is mainly responsible for the weaker splitting of the $^{3}H_{5}$
state.

Thus, despite of certain model inaccuracies which mainly concern the $A_{k}$s, the correct ratios of the
$\sigma^{2}$ of various states are available by calculation (Eq.(10)). These ratios can help us to choose the
proper parametrization of the $\mathcal{H}_{\mathrm{CF}}$, as well as to verify any considered one.

Finally, with the expression for $\sigma^{2}$ (Eq.(8)) we can estimate the adequacy of the
$S=(\sum_{k}S_{k}^{2})^{1/2}$ and $A=(\sum_{k}A_{k}^{2})^{1/2}$ concepts. Both these quantities are
simultaneously rational only in the three following cases:
\begin{itemize}
    \item [(i)] if all the three values of $|A_{k}|$ $(k=2,4,6)$ are close each other, i.e. $A_{k}^{2}\simeq
    \frac{1}{3}A^{2}$ [9,11]; then $\sigma^{2}(|\alpha SLJ\rangle)\simeq
    \frac{1}{2J+1}(S_{2}^{2}+S_{4}^{2}+S_{6}^{2})\frac{1}{3}A^{2}=\frac{1}{2J+1}\frac{1}{3}S^{2}A^{2}$,
    \item [(ii)] if all the three values of $S_{k}$ $(k=2,4,6)$ are close each other, i.e. $S_{k}^{2}\simeq
    \frac{1}{3}S^{2}$; then $\sigma^{2}(|\alpha SLJ\rangle)\simeq
    \frac{1}{2J+1}\frac{1}{3}S^{2}(A_{2}^{2}+A_{4}^{2}+A_{6}^{2})=\frac{1}{2J+1}\frac{1}{3}S^{2}A^{2}$,
    \item [(iii)] if there is a fully random distribution of the $S_{k}$ and $A_{k}$ values leading to the
    approximation $S_{2}^{2}A_{2}^{2}+S_{4}^{2}A_{4}^{2}+S_{6}^{2}A_{6}^{2}\simeq \frac{1}{3}
    (S_{2}^{2}+S_{4}^{2}+S_{6}^{2})(A_{2}^{2}+A_{4}^{2}+A_{6}^{2})=\frac{1}{3}S^{2}A^{2}$.
\end{itemize}
In all other cases the concepts of the total $S$ and $A$ lose the explicit link with the $\sigma^{2}$ and only
their conventional qualitative interpretation remains in power.

Concluding, the above analysis reveals the equivalent role of both the multipole distributions and throw the
light on the $S$ and $A$ physical meaning as well as on the limitations in their use.

\renewcommand{\baselinestretch}{1}
\begin{small}
\begin{table}[htbp]
\begin{center}
\caption{Multipole characteristics of the electron energy eigenstates $(^{2S+1}L_{J})$ coming from the ground
terms of $p^{n}$, $d^{n}$ and $f^{n}$ configurations (and similarly for $p^{6-n}$, $d^{10-n}$ and $f^{14-n}$
configurations)}
\vspace*{0.3cm}
\begin{tabular}{||c|c|l|r|r|r|r||}
\hline \hline
No. & Electron & Electron & $A_{2}=$ &$A_{4}=$ & $A_{6}=$ & $A=$\\
 & configuration & eigenstate &$\left<J||C^{(2)}||J\right>$ & $\left<J||C^{(4)}||J\right>$&
$\left<J||C^{(6)}||J\right>$ & $\left(A_{2}^{2}+A_{4}^{2}+A_{6}^{2}\right)^{1/2}$
\\
\hline \hline
 &  &  &  &   &  &\\
 1              & $p^{1}$           &     $^{2}{\rm P}$       &    -1.0955      &  0    &  0  & 1.0955\\
 2              &                   &     $^{2}{\rm P}_{3/2}$ &    -0.8944      &  0    &  0  & 0.8944\\
 3              &                   &     $^{2}{\rm P}_{1/2}$ &     0           &  0    &  0  & 0\\
 &  &  &  &   & &\\
\hline
&  &  &  &   &  &\\
 4              & $p^{2}$           &     $^{3}P$       &     1.0955      &  0    &  0  &1.0955\\
 5              &                   &     $^{3}P_{2}$   &     0.8367      &  0    &  0  &0.8367\\
 6              &                   &     $^{3}P_{1}$   &    -0.5477      &  0    &  0  &0.5477\\
 7              &                   &     $^{3}P_{0}$   &    0            &  0    &  0  &0\\
 &  &  &  &   & &\\
\hline
 &  &  &  &   & &\\
 8              & $d^{1}$           &     $^{2}D$       &    -1.1952      &  1.1952 &0 &1.6903\\
 9              &                   &     $^{2}D_{5/2}$       &    -1.1711      &  0.7559 &0 &1.3939\\
 10              &                   &    $^{2}D_{3/2}$       &    -0.8944      &  0 &0 &0.8944\\
 &  &  &  &   & &\\
\hline
 &  &  &  &   & &\\
 11              & $d^{2}$           &     $^{3}F$       &    -0.5855      &  -1.7728 &0 &1.8670\\
 12              &                   &     $^{3}F_{4}$       &    -0.6007      &  -1.3698 &0 &1.4957\\
 13              &                   &     $^{3}F_{3}$       &    -0.4392      &  -0.2955 &0 &0.5294\\
 14              &                   &     $^{3}F_{2}$       &    -0.4098      &  -0.6261 &0 &0.7483\\
 &  &  &  &   & &\\
\hline
 &  &  &  &   & &\\
 15              & $d^{3}$           &     $^{4}F$       &    0.5855      &  1.7728 &0 &1.8670\\
 16              &                   &     $^{4}F_{9/2}$       &    0.6117      &  1.2732 &0 &1.4125\\
 17              &                   &     $^{4}F_{7/2}$       &    0.3943      &  -0.1688 &0 &0.4289\\
 18              &                   &     $^{4}F_{5/2}$       &    0.2760      &  0.5940 &0 &0.6550\\
 19              &                   &     $^{4}F_{3/2}$       &    0.3067      &  0 &0 &0.3067\\
 &  &  &  &   & &\\
\hline
 &  &  &  &   & &\\
 20              & $d^{4}$           &     $^{5}D$       &    1.1952      &  -1.1952 &0 &1.6903\\
 21              &                   &     $^{5}D_{4}$       &   1.2014      &  -0.4565 &0 &1.2852\\
 22              &                   &     $^{5}D_{3}$       &    0.2928      & 0.8864 &0 &0.9335\\
 23              &                   &     $^{5}D_{2}$       &    -0.2561      &  -0.3415 &0 &0.4269\\
 24              &                   &     $^{5}D_{1}$       &    -0.4583      &  0 &0 &0.4583\\
 25              &                   &     $^{5}D_{0}$       &    0      &  0 &0 &0\\
 &  &  &  &   & &\\
\hline
 &  &  &  &   & &\\
 26              & $f^{1}$           &     $^{2}F$       &    -1.3663      &  1.1282 &-1.2774 &2.1843\\
 27              &                   &     $^{2}F_{7/2}$       &  -1.3801      & 0.9670 &-0.3054 &1.7126\\
 28              &                   &     $^{2}F_{5/2}$       &  -1.1711      &0.7560 &0 &1.3939\\
 &  &  &  &   & &\\
\hline
\end{tabular}
\end{center}
\end{table}
\newpage
\begin{table}[htbp]
\noindent Table 1 - cont.
\begin{center}
\begin{tabular}{||c|c|l|r|r|r|r||}
\hline \hline
No. & Electron & Energy & $A_{2}=$ &$A_{4}=$ & $A_{6}=$ & $A=$\\
 & configuration & eigenstate &$\left<J||C^{(2)}||J\right>$ & $\left<J||C^{(4)}||J\right>$&
$\left<J||C^{(6)}||J\right>$ & $\left(A_{2}^{2}+A_{4}^{2}+A_{6}^{2}\right)^{1/2}$
\\
\hline \hline
 &  &  &  &   &  &\\
 29              & $f^{2}$           &     $^{3}H$       &    -1.4555      &  1.0249 &1.4837 &2.3174\\
 30              &                   &     $^{3}H_{6}$       &    -1.5158      &  0.9583 &1.1386 &2.1242\\
 31              &                   &     $^{3}H_{5}$       &    -1.3100      &  0.6833 &0.4451 &1.5431\\
 32              &                   &     $^{3}H_{4}$       &    -1.2367     &  0.7395 &0.7706 &1.6340\\
 &  &  &  &   & &\\
\hline
&  &  &  &   & &\\
 33              & $f^{3}$           &     $^{4}I$       &    -0.6064      &  -0.7187 &-2.2771 &2.4636\\
 34              &                   &     $^{4}I_{15/2}$       &   -0.6438     & -0.6850 &-1.7999 &2.0306\\
 35              &                   &     $^{4}I_{13/2}$       &    -0.5569     &  -0.4691 &-0.6217 &0.9574\\
 36              &                   &     $^{4}I_{11/2}$       &   -0.5045     &  -0.3935 &-0.3399 &0.7245\\
 37              &                   &     $^{4}I_{9/2}$       &   -0.4954     &  -0.4904 &-1.1085 &1.3095\\
 &  &  &  &   & &\\
\hline
 &  &  &  &   & &\\
 38             & $f^{4}$            &     $^{5}I$       & 0.6064    &  0.7187 &2.2771 &2.4636\\
 39              &                   &     $^{5}I_{8}$       & 0.6562    &  0.6797 &1.7060 &1.9501\\
 40              &                   &     $^{5}I_{7}$       & 0.5524    &  0.4042 &0.2398 &0.7253\\
 41              &                   &     $^{5}I_{6}$       & 0.4796    &  0.2613 &-0.3105 &0.6283\\
 42              &                   &     $^{5}I_{5}$       & 0.4428    &  0.2437 &-0.2958 &0.5856\\
 43              &                   &     $^{5}I_{4}$       & 0.4540    &  0.4103 &0.7679 &0.9819\\
 &  &  &  &   & &\\
\hline
&  &  &  &   &  &\\
 44              & $f^{5}$           &     $^{6}H$       & 1.4555 &  1.0249 &-1.4837 &2.3174\\
 45              &                   &     $^{6}H_{15/2}$       & 1.6095 &  0.9134 &-0.9000 &2.0579\\
 46              &                   &     $^{6}H_{13/2}$       & 1.2375 &  0.3128 &0.4146 &1.3421\\
 47              &                   &     $^{6}H_{11/2}$       & 0.9587 &  -0.0108 &0.7152 &1.1961\\
 48              &                   &     $^{6}H_{9/2}$       & 0.7786 &  -0.1423 &0.6845 &1.0464\\
 49              &                   &     $^{6}H_{7/2}$       & 0.7176 &  -0.1129 &0.7035 &1.0112\\
 50              &                   &     $^{6}H_{5/2}$       & 0.8458 &  0.2978 &0 &0.8967\\
 &  &  &  &   & &\\
\hline
 &  &  &  &   & &\\
 51              & $f^{6}$           &     $^{7}F$              & 1.3663  & -1.1282 &1.2774 &2.1843\\
 52              &                   &     $^{7}F_{6}$              & 1.5159  & -0.7188 &0.2277 &1.6931\\
 53              &                   &     $^{7}F_{5}$              & 0.7277  & 0.5125 &-0.7419 &1.1587\\
 54              &                   &     $^{7}F_{4}$              & 0.1528  & 0.6075 &0.7650 &0.9888\\
 55              &                   &     $^{7}F_{3}$              & -0.2277  & 0.1880 &-0.2129 &0.3640\\
 56              &                   &     $^{7}F_{2}$              & -0.4382  & -0.3984 &0 &0.5922\\
 57              &                   &     $^{7}F_{1}$              & 0  & 0 &0 &0\\
 58              &                   &     $^{7}F_{0}$              & 0  & 0 &0 &0\\
\hline \hline
\end{tabular}
\end{center}
\end{table}
\end{small}

\renewcommand{\baselinestretch}{1}
\begin{table}[htb]
\begin{center}
\caption{Multipole characteristics of selected electron eigenstates $(^{2S+1}L_{J})$ distinguished by: the
strongest A (rows 1--10), very weak A (rows 11--20), domination$^{\ast}$ of $|A_{2}|$ (rows 21--33),
domination$^{\ast\ast}$ of $|A_{4}|$ (rows 34--41), domination$^{\ast\ast\ast}$ of $|A_{6}|$ (rows 42--44), the
comparable contribution of $|A_{2}|$, $|A_{4}|$, and $|A_{6}|$ (rows 45--56). In the last rows (57--63) there
are seven states $^{3}G_{5}$ coming from the seven various terms $^{3}G$ of $f^{6}$ configurations. $^{\ast}$
$\left|\frac{A_{2}}{A_{4}}\right|>5$ and $\left|\frac{A_{2}}{A_{6}}\right|>5$; \hspace*{0.1cm} $^{\ast\ast}$
$\left|\frac{A_{4}}{A_{2}}\right|>5$ and $\left|\frac{A_{4}}{A_{6}}\right|>5$; \hspace*{0.1cm} $^{\ast\ast\ast}$
$\left|\frac{A_{6}}{A_{2}}\right|>5$ and $\left|\frac{A_{6}}{A_{4}}\right|>5$}
\vspace*{0.1cm}
\begin{tabular}{||c|c|l|r|r|r|r||}
\hline \hline
No.  & Electron & Electron & $A_{2}=$ &$A_{4}=$ & $A_{6}=$ & $A=$\\
 & configuration & eigenstate &$\left<J||C^{(2)}||J\right>$ & $\left<J||C^{(4)}||J\right>$&
$\left<J||C^{(6)}||J\right>$ & $\left(A_{2}^{2}+A_{4}^{2}+A_{6}^{2}\right)^{1/2}$
\\
\hline \hline
 &  &  &  &   &  &\\
   1                &  $f^{4}$      &     $^{1}N$       & -3.5254  &  -1.7951 &1.4760 &4.2225\\
   2                &  $f^{3}$      &     $^{2}L_{17/2}$       &   -3.3430    &  -0.2254 &0.6523 &3.4136\\
   3                &  $f^{6}$      &    $^{1}Q$              & -1.5036 & -1.3591 & -2.7260 &3.3969\\
   4                &  $f^{2}$      &     $^{1}I$       &    -3.0318     &  1.4375 &-0.4554 &3.3861\\
   5                &  $f^{5}$      &     $^{2}O_{23/2}$       & -2.5911 & -1.5800 &-0.6924 &3.1128\\
   6                &  $f^{4}$      &    $^{3}M_{10}$       & -2.4678  &  0 &-1.6235 &2.9539\\
   7                &  $f^{4}$      &     $^{1}H(2)$       & -2.6199  &  -0.6833 &0.4451 &2.7439\\
   8                &  $f^{5}$      &     $^{2}K(4)_{15/2}$       & -2.4906 & 0.4092 &0.9639 &2.7018\\
   9                &  $d^{4}$      &    $^{1}I$       &   -1.2994      &  -2.2589 &0 &2.6060\\
  10                &  $d^{2}$      &     $^{1}G$       &    -2.4026      &  0.9131 &0 &2.5703\\
 &  &  &  &   & &\\
\hline
 &  &  &  &   &  &\\
  11                &  $f^{6}$      &     $^{3}G(1)_{4}$              & -0.0227 & -0.0296 &0.0099 &0.0386\\
  12                &  $f^{6}$       &    $^{3}H(2)_{5}$              & 0.0437 & 0.0228 &-0.0148 &0.0515\\
  13                &  $f^{6}$       &    $^{3}F(1)_{3}$              & 0.0683 & -0.0125 &-0.0639 &0.0944\\
  14                &  $f^{4}$       &    $^{3}G(1)_{4}$       & -0.0681    &  -0.0887 &0.0298 &0.1157\\
  15                &  $f^{5}$       &    $^{4}F(3)_{9/2}$       & -0.0793 &  -0.0696 &-0.0691 &0.1261\\
  16                &  $f^{5}$       &    $^{4}I(3)_{11/2}$       & 0.0721 & 0.1218 &0.0340 &0.1456\\
  17                &  $f^{6}$       &    $^{5}F(1)_{3}$              & -0.0721  & -0.0940 &-0.0887 &0.1480\\
  18                &  $f^{4}$       &    $^{3}H(2)_{5}$       & 0.1310  &  0.0683 &-0.0445 &0.1543\\
  19                &  $f^{6}$       &    $^{3}H(6)_{5}$              & -0.0621 & 0.1520 &-0.0163 &0.1650\\
  20                &  $f^{6}$       &    $^{3}G(2)_{4}$              & 0.1709 & 0.0059 &-0.0070 &0.1711\\
 &  &  &  &   & &\\
\hline
 &  &  &  &   &  &\\
  21                &  $f^{3}$       &    $^{2}L_{17/2}$       &   -3.3430    &  -0.2254 &0.6523 &3.4136\\
  22                &  $f^{3}$       &    $^{2}F(2)_{7/2}$       &  -1.6102   &  -0.1026 &0.0828 &1.6156\\
  23                &  $f^{4}$       &    $^{3}G(2)_{4}$       & 0.5128   &  0.0176 &-0.0211 &0.5135\\
  24                &  $f^{5}$       &    $^{4}H(2)_{11/2}$       & -0.3828 & 0.0079 &-0.0100 &0.3830\\
  25                &  $f^{5}$       &    $^{4}H(3)_{11/2}$       & 0.3904 & 0.0477 &-0.0302 &0.3945\\
  26                &  $f^{5}$       &    $^{2}L(1)_{17/2}$       & -1.6715 & -0.1127 &0.3262 &1.7068\\
  27                &  $f^{5}$       &    $^{2}N_{21/2}$       & -1.7598 & -0.0817 &-0.0786 &1.7634\\
  28                &  $f^{6}$       &    $^{5}H(2)_{6}$              & 1.1272  & -0.0184 &-0.1741 &1.1407\\
  29                &  $f^{6}$       &    $^{5}I(2)_{7}$              & -1.0521 & 0.0385 &-0.1279 &1.0605\\
  30                &  $f^{6}$       &    $^{3}F(6)_{4}$              & -0.6725 & -0.0487 &0.1009 &0.6818\\
  31                &  $f^{6}$       &    $^{3}G(2)_{4}$              & 0.1709 & 0.0059 &-0.0070 &0.1711\\
  32                &  $f^{6}$       &    $^{3}G(7)_{4}$              & -0.9812 & -0.0064 &0.0184 &0.9814\\
  33                &  $f^{6}$       &    $^{3}L(2)_{8}$              & -1.2705 & -0.1478 & 0.2320 &1.2999\\
 &  &  &  &   & &\\
\hline
\end{tabular}
\end{center}
\end{table}
\clearpage
\begin{table}[htb]
\noindent Table 2 - cont.
\begin{center}
\begin{tabular}{||c|c|l|r|r|r|r||}
\hline \hline
No.  & Electron & Electron & $A_{2}=$ &$A_{4}=$ & $A_{6}=$ & $A=$\\
 & configuration & eigenstate &$\left<J||C^{(2)}||J\right>$ & $\left<J||C^{(4)}||J\right>$&
$\left<J||C^{(6)}||J\right>$ & $\left(A_{2}^{2}+A_{4}^{2}+A_{6}^{2}\right)^{1/2}$
\\
\hline \hline
 &  &  &  &   &  &\\
   34                &  $f^{3}$      &    $^{2}H(2)_{11/2}$       &  -0.0076   &  0.5373 &0.0066 &0.5374\\
   35                &  $f^{3}$      &    $^{2}H(2)_{9/2}$       &  -0.0069   &  0.4816 &0.0057 &0.4817\\
   36                &  $f^{5}$      &    $^{2}H(2)_{11/2}$       & -0.0038 & 0.2686 &0.0033 &0.2686\\
   37                &  $f^{5}$      &    $^{2}H(2)_{9/2}$       & -0.0035 & 0.2408 &0.0029 &0.2408\\
   38                &  $f^{6}$      &    $^{3}G(4)_{4}$              & 0.0103 & 0.2824 &0.0326 &0.2845\\
   39                &  $f^{6}$      &    $^{3}G(5)_{5}$              & -0.0597 & 0.3596 &0.0065 &0.3646\\
   40                &  $f^{6}$      &    $^{3}G(5)_{3}$              & -0.0458 & 0.2451 &0.0029 &0.2494\\
   41                &  $f^{6}$      &    $^{1}N(2)$              & -0.2044 & 1.2618 & 0.1626 &1.2885\\
&  &  &  &   & &\\
\hline
 &  &  &  &   &  &\\
   42                &  $f^{6}$      &    $^{5}G(2)_{5}$              & -0.0491  & 0.0322 &0.4586 &0.4623\\
   43                &  $f^{6}$      &    $^{5}G(2)_{4}$              & -0.0362  & -0.0327 &0.4948 &0.4972\\
   44                &  $f^{6}$      &    $^{5}G(2)_{3}$              & -0.0302  & -0.0438 &0.4935 &0.4964\\
 &  &  &  &   & &\\
\hline
 &  &  &  &   &  &\\
   45                &  $f^{4}$      &    $^{3}G(3)$       & -0.8086   &  -0.7396 &0.7706 &1.3397\\
   46                &  $f^{6}$      &    $^{3}G(3)$              & -0.2696 & -0.2465 &0.2569 &0.4466\\
   47                &  $f^{1}$      &    $^{2}F$       &    -1.3663      &  1.1282 &-1.2774 &2.1843\\
   48                &  $f^{2}$      &    $^{3}F$       &    0.4554     &  -0.3761 &0.4258 &0.7281\\
   49                &  $f^{3}$      &    $^{4}F$       &   0.6831     &  -0.5641 &0.6387 &1.0921\\
   50                &  $f^{4}$      &    $^{5}F$       & -0.6831    &  0.5641 &-0.6387 &1.0921\\
   51                &  $f^{5}$      &    $^{6}F$       & -0.4554 &  0.3761 &-0.4258 &0.7281\\
   52                &  $f^{6}$      &    $^{7}F$              & 1.3663  & -1.1282 &1.2774 &2.1843\\
   53                &  $f^{6}$      &    $^{7}F_{3}$              & -0.2277  & 0.1880 &-0.2129 &0.3640\\
   54                &  $f^{5}$      &    $^{4}F(3)_{9/2}$       & -0.0793 &  -0.0696 &-0.0691 &0.1261\\
   55                &  $f^{4}$      &    $^{1}L(1)$       & -1.0938  &  -1.1329 &0.9100 &1.8188\\
   56                &  $f^{5}$      &    $^{4}K(2)_{11/2}$       & -0.5945 & -0.6108 &-0.7334 &1.1244\\
    &  &  &  &   & &\\
\hline
 &  &  &  &   &  &\\
   57                &  $f^{6}$      &   $^{3}G(1)_{5}$              & -0.0277 & -0.0521 &-0.1259 &0.1390\\
   58                &  $f^{6}$      &   $^{3}G(2)_{5}$              & 0.2089 & 0.0103 &0.0893 &0.2274\\
   59                &  $f^{6}$      &   $^{3}G(3)_{5}$              & -0.2799 & -0.2174 &0.1631 &0.3901\\
   60                &  $f^{6}$      &   $^{3}G(4)_{5}$              & 0.0126 & 0.4980 &-0.4137 &0.6475\\
   61                &  $f^{6}$      &   $^{3}G(5)_{5}$              & -0.0597 & 0.3596 &0.0065 &0.3646\\
   62                &  $f^{6}$      &   $^{3}G(6)_{5}$              & 0.3602 & 0.2853 &-0.2503 &0.5232\\
   63                &  $f^{6}$      &   $^{3}G(7)_{5}$              & -1.1988 & -0.0112 &-0.2335 &1.2214\\
&  &  &  &   & &\\
\hline \hline
\end{tabular}
\end{center}
\end{table}

\clearpage



\end{document}